\documentstyle[11pt]{article}

  \let\bsk=\bigskip
\let\qd=\quad  
\let\a=\alpha \let\be=\beta \let\g=\gamma \let\de=\delta
\let\ep=\varepsilon   
  \let\la=\lambda 
   \let\si=\sigma
 
\let\ph=\varphi \let\Ph=\phi \let\PH=\Phi \let\Ps=\Psi
 \let\Si=\Sigma 
 \let\Ga=\Gamma 
\def\da{{\dot \a}} \def\dbe{{\dot \be}}

\def\0#1#2{\frac{#1}{#2}} \def\s0#1#2{\mbox{\small{$\frac{#1}{#2}$}}}
\def\2{{\times}} \def\3{\vec }
\def\5{\bar }  \def\6{\partial } \def\7{\hat } \def\4{\tilde }
 \def\lb{\left(} \def\rb{\right)}
 \let\LRA=\Leftrightarrow
\let\then=\Rightarrow 
\def\sl2c{$sl(2,{\bf C})$}
\def\SL2c{$SL(2,{\bf C})$}

\let\nn=\nonumber
\def\bea{\begin{eqnarray}} \def\eea{\end{eqnarray}}
\def\beann{\begin{eqnarray*}} \def\eeann{\end{eqnarray*}}
\def\beq{\begin{equation}} \def\eeq{\end{equation}}
\def\ba{\begin{array}} \def\ea{\end{array}}
\def\ben{\begin{enumerate}} \def\een{\end{enumerate}}

 \def\cb{{\cal B}} \def\cg{{\cal G}}
\def\ca{{\cal A}} \def\cd{{\cal D}} \def\cl{{\cal L}}
  
\def\cF{{\cal F}}  
\def\cW{{\cal W}}

\def\F#1#2#3{{F_{#1#2}}^{#3}}

\def\A#1#2{{A_{#1}}^{#2}}

\newcommand{\mysection}[1]{\section{#1}
            \setcounter{equation}{0}\setcounter{figure}{0}}
\def\Gl#1{(\ref{#1})}
\def\act{S} \def\ano{\cW^1} \def\cla{S_{cl}} \def\ppp{[\PH,\PH^*]}
\def\brs{\cb_\act}
\def\gao{\Ga^{(0)}}
\def\bgao{\cb_{\gao}}
\def\anoo{\ca^{(\ell)}}
\def\an#1{\cW^1_{#1}}

\def\pp{\PH,\PH^*}
\def\LM{b}
\def\bb#1{\cb_{#1}}

\def\PPs{{\overline \Ps}}
\begin{document}
{\pagestyle{empty}
\vspace*{2cm}

\hspace*{\fill} NIKHEF-H 93--22\vspace*{1cm}\\
\begin{center}{\LARGE {\bf Antifield Dependence of Anomalies}}\vspace{1cm}

{\renewcommand{\thefootnote}{\fnsymbol{footnote}}
{\Large Friedemann Brandt}\footnote{Supported by Deutsche
Forschungsgemeinschaft}}
\setcounter{footnote}{0}\vspace{.5cm}

NIKHEF-H, Postbus 41882,
1009 DB Amsterdam, The Netherlands\end{center}\vspace{.5cm}

\begin{abstract}
It is shown that generally the consistency equation for anomalies
of quantum field theories has
solutions which depend nontrivially on the sources of the
(generalized) BRS-transformations of the fields. Explicit
previously unknown examples of such solutions
are given for Yang-Mills and super Yang-Mills theories.
\end{abstract}
\newpage}\setcounter{page}{1}
\mysection{Introduction}\label{intro}

Gauge symmetries are of central importance in our present theories of
fundamental interactions. If there is no
regularization procedure which respects all symmetries of
the classical theory it is not guaranteed that
these symmetries survive the quantization of the theory, i.e.
the theory may turn out to be anomalous.
The BRS-formalism \cite{brs,bau1} allows to
characterize anomalies as solutions of the so-called
consistency equation which generalizes the Wess-Zumino consistency
conditions \cite{wz}. This makes
possible an algebraic classification of anomaly candidates without
referring to a particular regularization scheme.
A generalization of the BRS-formalism to symmetries whose algebra
does not necessarily close off-shell is given by the BV-antifield-formalism
\cite{bv1} whose forerunner has been formulated in \cite{hw}.
Surprisingly it took quite a long time until anomalies have been discussed
in this formalism \cite{tro} although the BV-formulation allows to
characterize anomalies analogously as solutions of a consistency equation
which follows from the anomalous Slavnov--Taylor--Ward identity
{}for the effective action \cite{how}.

The consistency equation is most conveniently written in terms
of the fields and antifields and it can be shown that each of its
nontrivial solutions contains an antifield independent part
which characterizes and determines it almost completely
\cite{proe}. Nevertheless in general the complete
solution depends on the antifields resp.
on the sources of the (generalized) BRS-transformations of the fields.
The main point I want to make in this paper is to show that this
dependence is generally nontrivial,
whether the algebra of the classical symmetries closes off-shell or not.
Namely by means of an explicit example I show that even in the simple and
phenomenologically important case of
four dimensional renormalizable Yang-Mills theories
there are previously unknown solutions of the
consistency equation which depend nontrivially on the sources (antifields),
contrary to the common belief and to different and consequently
erroneous statements which have been given in the literature \cite{bau2}.

Furthermore I discuss the
connection between the BV- and the usual BRS-formu\-la\-tion of
theories with symmetries
whose algebra can be closed off-shell by means of appropriate auxiliary fields.
In particular it is shown that solutions of the consistency
equation which do not depend on the antifields in the
formulation with auxiliary fields will generally
(but not necessarily) depend on the antifields
in the formulation without auxiliary fields.
This result is exemplified
{}for the case of abelian super-Yang-Mills theories where it yields alternative
forms of recently found new anomaly candidates of these theories.

\mysection{The consistency equation}\label{cc}

I shall now briefly discuss the consistency equation in the BV-formalism.
An anomaly shows up as a violation of the Slavnov-Taylor-Ward identity for the
generating functional $\Ga[\PH,q,\zeta,\LM]$ of renormalized 1PI Green
functions
which depends on the classical fields and the ghosts, denoted collectively
by $\PH$, as well as on the sources $q$ of their generalized
BRS-transformations, the antighosts $\zeta$ and the Lagrange multiplier
fields $\LM$. $\Ga$ is constructed order by order in a loop expansion
\beq \Ga=\sum_{n\geq 0}\hbar^n\Ga^{(n)}\label{c3}\eeq
where the tree functional $\gao$ contains the invariant classical
action as well as gauge fixing and corresponding ghost contributions
(see below). It is chosen such that $\Ga$ satisfies
the Slavnov-Taylor identity in 0th order:
\beq \bgao\gao=0\label{c2a}\eeq
where $\bgao$ is the nilpotent operator
\beq \bgao=\int d^D\! x  \, \lb \0 {\de \gao}{\de q_A}
\0 {\de}{\de \PH^A}
+\0 {\de \gao}{\de \PH^A}\0 {\de}{\de q_A}+\LM^N\0 {\de}{\de \zeta^N}\rb.
\label{c8}\eeq
In the anomalous case \Gl{c2a} cannot be extended to all orders
and the Slavnov-Taylor identity for $\Ga$ is violated by an anomaly $\ca$
occurring at some order $\ell$ which is nonzero
due to \Gl{c2a} but generally not known in advance:
\beq \cb_\Ga\Ga=\ca=\sum_{n\geq \ell}\hbar^n\ca^{(n)},\qd\ell>0. \label{c1}\eeq
The lowest order contribution
$\anoo$ is a local functional and satisfies the consistency equation
\beq \bgao \anoo =0\label{c7}\eeq
which follows from
the identity $\cb_\Ga\cb_\Ga\Ga=0$ since the latter and \Gl{c1} imply
$\cb_\Ga\ca=0$ whose lowest order contribution is just \Gl{c7}.
Trivial contributions
$\bgao X$ can be removed from $\anoo$ by subtracting the counterterm
$\hbar^\ell X$ from $\Ga$.
In particular $\anoo$ itself can be assumed to be nontrivial since
otherwise the anomaly can be removed up to terms of order $n>\ell$.

In the BV-formalism $\gao$ is constructed from the proper solution $\act$
of the so-called classical master equation which has the form
\beq \act=\act\ppp=\cla[\Ph]+\int d^D\! x\, \PH^*_AR^A(\PH)+O(2)\label{z1}\eeq
where $O(2)$ collects all terms which are at least bilinear in the
antifields $\PH^*$, $\cla$ is the classical invariant action and the
$R^i$ generate its symmetries.
Here $\{\PH^A,\PH^*_A\}$ denote collectively the
minimal set of fields and antifields in the sense of \cite{bv1} which consists
of the classical fields
$\Ph^i$, the ghosts $C^N$ and their respective
antifields\footnote{For simplicity only gauge theories are considered
which are irreducible in the sense of \cite{bv1} though everything extends
straightforwardly to the reducible case as well.}:
\beq \{\PH^A\}=\{\Ph^i,C^N\},\qd \{\PH^*_A\}=\{\Ph^*_i,C^*_N\}.\label{i4}\eeq
In order to construct $\gao$ one first adds
the term $\zeta^*_N\LM^N$ to the integrand of $\act$ and then one
fixes the gauge by means of an appropriate fermionic functional
$\Ps[\PH,\zeta,\LM]$ with ghost number $-1$:
\bea & &\left.\gao=\lb\act\ppp+\int d^D\! x  \, \zeta^*_N\LM^N
\rb\right|_{\Si'},\label{c10}\\
& &{\Si'}:\qd \PH^*_A=q_A-(-)^{\ep(\PH^A)}\0 {\de\Ps}{\de\PH^A},\qd
\zeta^*_N=-(-)^{\ep(\zeta^N)}\0 {\de\Ps}{\de\zeta^N}
\label{c11}\eea
where $\ep(Z)$ denotes the grading of $Z$ (the signs occur since I use
leftderivatives only). Straightforwardly
one shows by means of standard methods that solutions
of \Gl{c7} depend on the fields $\zeta^N$, $\LM^N$ only trivially if
written in terms of the variables $\{\PH^A,\PH^*_A,\zeta^N,\LM^N\}$
since in these variables the generalized BRS-transformations take
the simple form
\[ \bgao\PH^A=\brs \PH^A,\qd  \bgao\PH^*_A=\brs \PH^*_A,\qd
\bgao\zeta^N=\LM^N,\qd \bgao \LM^N=0\]
where $\brs$ is the operator
\beq \brs =\int d^D\! x\,  \lb \0 {\de \act}{\de \PH^*_A}
\0 {\de}{\de \PH^A}
+\0 {\de\act}{\de \PH^A}\0 {\de}{\de \PH^*_A}\rb.
\label{i2}\eeq
As a result we obtain
\beq \anoo=\ano\ppp|_{\Si'}+\bgao X[\PH,q,\zeta,\LM]\label{c13}\eeq
where $X$ is a local functional with ghost number 0 and $\ano$ solves
\beq \brs \ano \ppp=0\label{i1}\eeq
which is the form of the consistency equation discussed in \cite{proe}.
Thus \Gl{c7} reduces to \Gl{i1} since $\bgao X$ in \Gl{c13} is a trivial
contribution to $\anoo$. Furthermore contributions
$\brs Y\ppp$ to solutions of \Gl{i1} obviously correspond to trivial
contributions $\bgao \7Y$ to $\anoo$ where $\7Y=Y|_{\Si'}$
and solve \Gl{i1} since
the master equation implies $(\brs)^2=0$. Therefore
two solution of \Gl{i1} are
called equivalent if they differ by such trivial contributions:
\beq \4\ano\ppp\cong\ano\ppp\ \LRA\ \4\ano\ppp-\ano\ppp=\brs Y\ppp.\label{i1a}
\eeq
It can be shown \cite{proe} that each
nontrivial solution of \Gl{i1} has a nonvanishing antifield independent
part $\an 0[\PH]$ which satisfies
\beq\bb 1\,\an 0[\PH]\sim 0,\qd\an 0[\PH]\not\sim\bb 1X_0[\Ph]\label{ns}\eeq
where $\bb 1$ is an operator whose action on the $\PH^A$ is defined by means
of the part of \Gl{z1} which is linear in the antifields,
\beq \bb 1\PH^A=R^A(\PH),\label{bb1}\eeq
and $\sim$ denotes `weak equality' defined according to
\beq \cF[\PH]\sim \cg[\PH]\qd :\LRA\qd \cF[\PH]-\cg[\PH]=\int d^D\! x\,
\0 {\de\cla[\Ph]}{\de\Ph^i}\,  Z^i(\PH)\label{sim}\eeq
where $Z^i(\PH)$ are arbitrary local functions of the $\PH^A$ and their
derivatives. Notice that \Gl{ns} imposes only on-shell conditions
on $\an 0$ since $\sim$ requires equality up to
contributions which contain the classical equations of motion.

The importance of \Gl{ns} consists in the fact that it represents a
necessary and sufficient condition for the existence and nontriviality
of the complete solution of \Gl{i1}. Namely
each solution of \Gl{ns} can be completed to a nontrivial solution
\beq \ano\ppp=\an 0[\PH]+O(1)\label{i3b}\eeq
of \Gl{i1} and each nontrivial solution of \Gl{i1} contains a solution
$\an 0[\PH]$ of \Gl{ns} \cite{proe}. This can be proved by means of a result
about the cohomology of the so-called Koszul--Tate differential which
holds under appropriate assumptions about the classical
action and the gauge transformations \cite{hen}.

\mysection{Symmetries whose algebra closes off-shell}
\label{off}

It is well-known that
the BV-formalism reduces to the usual BRS-formalism
{}for theories with symmetries whose algebra closes off-shell.
Let us briefly recall this fact. In the case of an off-shell
closing algebra one can define a BRS-operator $s$
 which is off-shell nilpotent on the classical fields and the ghosts.
The solution of the master equation then takes the simple form
\beq \act=\cla[\Ph]+\int d^D\! x\,    \PH^*_A\, s\PH^A\label{b1}\eeq
and \Gl{c10} can be written in the familiar form
\beq \gao= \cla [\Ph]+s\,\Ps[\PH,\zeta,\LM]+\int d^D\! x\,     q_A\, s \PH^A
\label{c4}\eeq
where $s$ acts on $\zeta^N$ and $\LM^N$ according to
$s\zeta^N=\LM^N$, $s\LM^N=0$. Due to \Gl{c4}
both $\brs\PH^A$ and $\bb 1\PH^A$
agree with the usual nilpotent BRS-transformations:
\beq \brs\PH^A=\bb 1\PH^A=s\PH^A,\qd s^2\PH^A=0.\label{b5a}\eeq
This implies in particular that each BRS-invariant
functional $\cF[\PH]$ with ghost number 1 solves \Gl{i1}
and thus represents an anomaly candidate which does not
depend on the antifields resp. on the sources of the BRS-transformations
of the fields:
\beq s\, \ano[\PH]=0\qd \LRA\qd \brs\ano[\PH]=0.\label{b6a}\eeq
\Gl{b6a} has been investigated extensively for various theories in the
literature. Complete results have been derived for instance in \cite{com}
{}for Yang-Mills and Einstein-Yang-Mills theories and in
\cite{dok,glusy,sugra} for a class of globally and locally supersymmetric
theories by means of methods which
can be generalized to a large class of gauge theories \cite{ten}.

However it is still an open question in which cases the solutions
of \Gl{b6a} cover already the complete space of solutions of
\Gl{i1}. In order to show that this is generally not the case
I give an explicit example of a source dependent solution
of \Gl{i1} in Yang--Mills theory
which is not equivalent to a solution of \Gl{b6a}.
\bsk

\noindent {\it Example:} I consider a
four dimensional renormalizable abelian Yang--Mills theory defined by
the following integrand of a solution of the master equation:
\bea \cl=\sum_{I} (-\s0 14\,\F abIF^{abI}+A^{*a}_I\6_aC^I)
+i\sum_j\PPs^j\g^a\cd_a\Ps^j
+\sum_{jI}C^I(\Ps^*_j\de_I\Ps^j
+\PPs^*_j\de_I\PPs^j)\label{bs1}\eea
where $\F abI=\6_a\A bI-\6_b\A aI$
are the abelian field strengths, $C^I$ are the abelian ghost fields and
$\{\Ps^j\}$ is a set of fermions in Dirac bi-spinor notation
($\PPs=\Ps^\dagger\g^0$).
$\de_I$ denotes the generator of the $I$th $U(1)$-factor
and $\cd_a$ denotes the covariant derivatives:
\[  \de_I\Ps^j=ig_I^j\Ps^j,\qd \de_I\PPs^j=-ig_I^j\PPs^j,\qd
\cd_a=\6_a-\sum_I\A aI\de_I\]
where $g_I^j$ is the charge of $\Ps^j$ under $\de_I$.
One may check that the following functional solves \Gl{i1} and is
not equivalent to a solution of \Gl{b6a}:
\beq \ano=\sum_{jIJ}k_{IJ}\int d^4\! x\, \lb C^I\A aJ\PPs^j\g^a\g_5\Ps^j
+\s0 i2\, C^IC^J(\Ps^*_j\g_5\Ps^j+\PPs^j\g_5\PPs^*_j)\rb\label{bs2}\eeq
where $k_{IJ}$ are antisymmetric constants:
\beq k_{IJ}=-k_{JI}.\label{bs3}\eeq
Of course the example can be extended by coupling the $\Ps$ to nonabelian
gauge fields as well.
Notice that due to \Gl{bs3} the anomaly candidates \Gl{bs2} occur
only if the gauge group contains at least two abelian factors.
I remark that the nontrivial dependence of \Gl{bs2}
on the antifields originates in the occurrence of $\g_5$ in
\Gl{bs2}.
Since $\cl$ does not contain $\g_5$-dependent contributions one
of course does not expect the presence of an anomaly which corresponds to
\Gl{bs2} in this simple model but it is not excluded that similar anomaly
candidates exist in more complicated theories.

\mysection{Elimination of auxiliary fields}\label{aux}

Often one can close an only on-shell closing algebra
also off-shell by means of an appropriate
set of auxiliary fields. However on the
one hand auxiliary fields enlarge the field content
unnecessarily in the BV-formalism and on the other hand it is
in practice often difficult to find a set of auxiliary fields.
Therefore it is instructive to compare the formulations of a theory
with and without auxiliary fields. To this end we denote by $\PH^A$
the classical fields and ghosts
which occur in the BV-formulation without auxiliary fields and
denote the latter by $H^r$. In the formulation with auxiliary fields
the solution of the master equation has the form \Gl{b1}:
\beq \7\act[\pp,H,H^*]=\7\cla[\PH,H]+\int d^D\! x\, (\PH^*_A\, s\PH^A+H^*_r\,
sH^r)
\label{a1}\eeq
where some of the nilpotent BRS-transformations $s\PH^A$ of course depend
on the auxiliary fields. As the defining property of the
auxiliary fields we require that the `equations of motion' for the $H^r$
which follow from \Gl{a1} after setting to zero $H^*_r$ have an
algebraic solution $\7H^r(\pp)$:
\beq 0=\0 {\de\7\act[\pp,H,0]}{\de H^r}\qd \LRA\qd
H^r=\7H^r(\pp).\label{a2}\eeq
Notice that this definition of auxiliary fields differs slightly from the
usual one which requires only that $\de\7\cla[\PH,H]/\de H^r$ can be
solved algebraically for the $H^r$. Our
definition is motivated by the fact that the elimination of
the auxiliary fields from \Gl{a1} by means of \Gl{a2} provides
directly the BV-formulation of the theory since
\beq \act\ppp:=\7\act[\pp,\7H(\pp),0]\label{a4}\eeq
solves the master equation. This result is contained in
the following lemma:
\bsk

\noindent {\it Auxiliary lemma:}
If a local functional $\7\cF[\pp,H,H^*]$ is $\cb_{\7\act}$-invariant
then the functional which arises from it for $H^r=\7H^r(\pp)$, $H^*_r=0$
is $\brs$-invariant:
\beq \cb_{\7\act}\7\cF[\pp,H,H^*]=0\qd\then\qd
\brs\7\cF[\pp,\7H(\pp),0]=0 \label{a6}\eeq
where $\brs$ denotes the operator \Gl{i2} arising from \Gl{a4} and
$\cb_{\7\act}$ denotes the analogous operator arising from \Gl{a1}
($\cb_{\7\act}$ contains functional derivatives with
respect to $H^r$ and $H^*_r$).
\bsk

{\it Proof:} In order to prove this lemma we introduce the notation
\[ G[\pp,H,H^*]|:=G[\pp,\7H(\pp),0]\]
where $G$ denotes an arbitrary functional of the $\pp$,$H$,$H^*$.
\Gl{a6} is proved as follows:
\beann\lefteqn{0=\int d^D\! x\,  \left.\lb (\cb_{\7\act}\PH^A)\0 {\de
\7\cF}{\de \PH^A}
+(\cb_{\7\act}\PH^*_A)\0 {\de \7\cF}{\de \PH^*_A}
+(\cb_{\7\act}H^r)\0 {\de \7\cF}{\de H^r}
+(\cb_{\7\act}H^*_r)\0 {\de \7\cF}{\de H^*_r}\rb\right|}\\
&&=\int d^D\! x\,  \left.\lb(\brs \PH^A)\0 {\de \7\cF}{\de \PH^A}
+(\brs\PH^*_A)\0 {\de \7\cF}{\de \PH^*_A}
+(\brs \7H^r)\0 {\de \7\cF}{\de H^r}\rb\right|=\brs (\7\cF|)\eeann
where the first row is obtained from writing out
$\cb_{\7\act}\7\cF[\pp,H,H^*]=0$ explicitly
and the second row follows from the first due to
\beq  (\cb_{\7\act}\PH^A)|=\brs \PH^A,\qd
(\cb_{\7\act}\PH^*_A)|=\brs \PH^*_A
,\qd (\cb_{\7\act}H^r)|=\brs H^r(\pp),
\qd (\cb_{\7\act}H^*_r)|=0 \label{brs}\eeq
which hold due to \Gl{a2}.\hfill $\Box$
\bsk

As mentioned above, the lemma implies in particular that \Gl{a4} solves
the master equation since $\7\act$ is $\cb_{\7\act}$-invariant by assumption
and the master equation can be written in the form
\beq \brs\act=0.\label{i3}\eeq
\Gl{a4} and \Gl{brs} are often useful in themselves since they facilitate
the construction of the solution of the master equation and the
generalized BRS-transformations considerably if
a formulation of the theory with auxiliary fields is known.
{}For instance by means of \Gl{a4} and \Gl{brs} one can easily
reproduce the results given in \cite{bau3} for N=1, D=4 supergravity
and super-Yang-Mills theories
(analogously to the example discussed below).
\bsk

\noindent {\it Example:} The outlined procedure is now
exemplified for abelian
D=4, N=1 super-Yang-Mills theories whose classical Lagrangian
reads in the formulation with auxiliary fields
\bea \7\cl_{cl} &=&\sum_{I}
(-\s0 14\, \F abIF^{abI}-i\,\la^I\si^a\6_a\5\la^I+\s0 12\, D^ID^I)
\nn\\ & &
+\sum_{j}(-\5\ph^j\cd_a\cd^a\ph^j-i\, \chi^j\si^a\cd_a\5\chi^j+ F^j\5F^j)
+V(\ph,\chi,F,\5\ph,\5\chi,\5F)
\nn\\
& &+\sum_{jI}(-iD^I\5\ph^j\de_I\ph^j+\sqrt 2\, \la^I\chi^j\de_I\5\ph^j+
\sqrt 2\, \5\la^I\5\chi^j\de_I\ph^j)\label{s1}\eea
where $\F abI$ denote as in the previous section
abelian field strengths, $\la_\a^I$ denote the gauginos,
$D^I$ are the real auxiliary fields of the super-Yang-Mills multipletts,
$\ph^j$ and $\chi^j_\a$ are component fields of chiral matter multipletts
and $F^j$ are the corresponding complex auxiliary fields.
Contrary to the previous section, a two component Weyl spinor
notation is used in \Gl{s1} in which $\la$ and $\5\la$ are related
just by complex conjugation\footnote{The conventions are the
same as in \cite{glusy}.}. Again
$\de_I$ denotes the generator of the $I$th $U(1)$-factor
and $\cd_a$ denotes the covariant derivatives
\[  \de_I\ph^j=ig_I^j\ph^j,\qd \de_I\5\ph^j=-ig_I^j\5\ph^j,\qd
\cd_a=\6_a-\sum_I\A aI\de_I\]
(the charges of $\chi^j$ and $F^j$ are also given by $g_I^j$).
$V$ denotes
contributions obtained from a $\de_I$-invariant superpotential $f(\ph)$
according to
\beq  V(\ph,\chi,F,\5\ph,\5\chi,\5F)=\cd^\a\cd_\a f(\ph)+
\5\cd_\da\5\cd^\da \5f(\5\ph)\label{sp1}\eeq
where $\cd_\a$ and $\5\cd_\da$ are spinor transformations defined by
\beq\ba{lll}\cd_\a\ph^j=\sqrt 2\, \chi^j_\a,&
 \cd_\a\chi^j_\be=\sqrt 2\, \ep_{\be\a}F^j,&
\cd_\a F^j=0,\\
\5\cd_\da\5\ph^j=\sqrt 2\, \5\chi^j_\da,&
 \5\cd_\da\5\chi^j_\dbe=\sqrt 2\, \ep_{\da\dbe}\5F^j,&
\5\cd_\da \5F^j=0.\ea
\label{sp2}\eeq
The nilpotent BRS-transformations under which $\7\cla=
\int d^4\! x\7\cl_{cl}$ is invariant
read
\bea s\A aI&=&\6_aC^I+i\la^I\si\5\xi-i\xi\si_a\5\la^I+C^b\6_b\A aI,
                                                             \label{s2a}\\
s\la_\a^I&=&-i\xi_\a D^I+\si^{ab}{}_\a{}^\be\xi_\be\F abI+C^a\6_a\la_\a^I,
                                                             \label{s2b}\\
sD^I&=&\6_a\la^I\si^a\5\xi+\xi\si^a\6_a\5\la^I+C^a\6_aD^I,
                                                             \label{s2c}\\
s\ph^j&=&\sqrt 2\, \xi\chi^j+C^I\de_I\ph^j+C^a\6_a\ph^j,
                                                             \label{s2d}\\
s\chi^j_\a&=&\sqrt 2\, \xi_\a F^j-\sqrt 2\,  i\5\xi^\da\cd_{\a\da}\ph^j
+C^I\de_I\chi^j_\a+C^a\6_a\chi^j_\a,
                                                             \label{s2e}\\
sF^j&=&-\sqrt 2\,  i\cd_a\chi^j\si^a\5\xi+2\5\xi\5\la^I\de_I\ph^j
+C^I\de_IF^j+C^a\6_aF^j,
                                                             \label{s2f}\\
sC^I&=&-2i\xi\si^a\5\xi\A aI+C^a\6_aC^I,
                                                             \label{s2g}\\
sC^a&=&2i\xi\si^a\5\xi,
                                                             \label{s2h}\\
s\xi^\a&=&0
                                                             \label{s2i}\eea
where $C^I$ are anticommuting Yang-Mills ghosts, $C^a$ denote constant
anticommuting ghosts of translations and $\xi^\a$, $\5\xi^\da$ are constant
commuting supersymmetry ghosts.
Nontrivial solutions of \Gl{b6a} are given by
\bea \7\ano_{chir}&=&\sum_{IJK}d_{IJK}\int d^4\! x\,  \lb
\ep^{abcd}\{ C^I \F abJ\F cdK+2i\A aI \F bcJ[\xi\si_d\5\la^K-
\la^K\si_d\5\xi]\}\right.\nn\\
& &\phantom{\sum_{IJK}d_{IJK}\int d^4\! x\,  \lb\rb.}\left.
+3i\,\{\xi\la^I \5\la^J\5\la^K+\5\xi\5\la^I \la^J\la^K\}\rb,
                                                            \label{d13}\\
 \7\ano_{fi}&=&\sum_{IJ} k_{IJ}\int d^4\! x\,
      \lb C^ID^J+\xi\si^a\5\la^I\A aJ+\la^I\si^a\5\xi\A aJ\rb  \label{s3}\eea
where the coefficients $d_{IJK}$ in \Gl{d13} are totally symmetric
and the coefficients $k_{IJ}$ in \Gl{s3} are antisymmetric:
\beq   d_{IJK}=d_{(IJK)},\qd k_{IJ}=-k_{JI}.\label{s3a}\eeq
\Gl{d13} are supersymmetric versions of abelian
chiral anomalies which have been derived in this or a similar form
{}for instance in \cite{kai,dok,sugra}. The solutions \Gl{s3} have been
found in \cite{dok}.
Notice that as the example of the previous section
their presence requires at least
two abelian factors due to the antisymmetry of $k_{IJ}$.

By means of the procedure outlined above one easily
constructs the BV-formu\-la\-tion of the theory without antifields. To this
end one eliminates the auxiliary fields $D^I$ and $F^j$ according to
\Gl{a2} which yields in this case:
\bea \7D^I=\sum_ji\,\5\ph^j\de_I\ph^j+ i\,\xi\la^*_I+i\,\5\xi\5\la^*_I,\qd
\7F^j=4\,\0 {\6\5f(\5\ph)}{\6\5\ph^j}+\sqrt 2\, \5\xi\5\chi^*_j\label{s4}\eea
where $\la^*_{I\a}$ and $\5\la^*{}^\da_I$ are the antifields of $\la^{I\a}$
and $\5\la^I_\da$
($\la^*{}^\a_I$ and $\5\la^*{}^\da_I$ are related by complex conjugation)
and $\5\chi^*{}^\da_j$ is the antifield of $\5\chi^j_\da$.
Notice that the chiral anomalies \Gl{d13} do not depend on
auxiliary fields at all. They therefore keep their form
and in particular do not depend on antifields even
in the BV-formulation without auxiliary fields.
This is different in the case of the anomaly
candidates \Gl{s3} since they depend on the auxiliary fields $D^I$ and
thus give rise to the following
antifield dependent solution of \Gl{i1} in the formulation without
auxiliary fields:
\beq \ano_{fi}=\int d^4\! x\,   \sum_{IJ} k_{IJ}
( \xi\si^a\5\la^I\A aJ+\la^I\si^a\5\xi\A aJ+i\, C^I\sum_j\5\ph^j\de_J\ph^j
+i\, C^I\xi\la^*_J+i\, C^I\5\xi\5\la^*_J).\label{s5}\eeq
One can check the nontriviality of \Gl{s5} by making shure that there
is no local counterterm $X[\Ph]$ such that
$\bb 1X\sim\ano_{fi}|_{\la^*=\5\la^*=0}$.
I remark that both \Gl{d13} and \Gl{s3} have locally supersymmetric
extensions which therefore provide anomaly candidates
of supergravity \cite{sugra}.

\end{document}